\documentclass[conference]{IEEEtran}
\IEEEoverridecommandlockouts   

\usepackage{amsmath,amssymb}
\usepackage{graphicx}
\usepackage{booktabs}
\usepackage{array}
\usepackage{multirow}
\usepackage{url}
\usepackage[draft]{hyperref}
\usepackage{tikz}
\usetikzlibrary{shapes.geometric, arrows.meta, positioning, fit, backgrounds}
\usepackage{cite}
\usepackage{xcolor}
\usepackage{colortbl}

\usepackage[letterpaper, left=0.625in, right=0.625in, top=0.75in, bottom=1.1in]{geometry}
\usepackage{fontspec}
\usepackage{xeCJK}
\usepackage{makecell}

\newcommand{\ipa}[1]{#1}
\definecolor{todo}{RGB}{200,0,0}

\begin{document}
\author{
  \IEEEauthorblockN{Diane Myung-kyung Woodbridge}
  \IEEEauthorblockA{Masters in Data Science and Artificial Intelligence Program\\ 
  University of San Francisco\\
  dwoodbridge@usfca.edu}
  \and
  \IEEEauthorblockN{Jee Hyun Suh}
  \IEEEauthorblockA{Department of Rehabilitation Medicine\\ 
  Seoul National University Bundang Hospital\\
  jeehyun.suh1@snubh.org}
}
\title{Automated Pronunciation Evaluation for Korean Toddler Speech using Speech Diarization and Self-Supervised Learning}

\maketitle
\vspace{-158pt}
\begin{abstract}
Speech sound disorders affect approximately 44\% of Korean pediatric communication disorder cases, yet automated assessment tools for Korean toddler speech remain underdeveloped. This paper presents an end-to-end pipeline for automated pronunciation evaluation of Korean toddler speech, combining neural speaker diarization with self-supervised speech representation learning. We introduce a novel IRB-approved corpus of 53 recordings from Korean-speaking children aged 2-5 years. A subset of 53 subjects was annotated by three independent reviewers, yielding 1,190 consonant and 748 vowel word-level binary correctness labels. We evaluate three  diarization models, finding that NeMo SortFormer achieves 88.69\% speaker count accuracy and 33.04\% diarization error rate (DER) owing to its arrival-time-sorted transformer architecture, which handles the acoustic confound between young female caregivers exhibiting aegyo and toddler speech. For pronunciation scoring, we compare three self-supervised learning (SSL) backbones across multiple pooling strategies. A cross-model ensemble routing consonant prediction to HuBERT-large and vowel prediction to WavLM-large achieves balanced accuracies of 0.720 and 0.845, with a mean of 0.782.
\end{abstract}

\begin{IEEEkeywords}
Pediatric Speech Evaluation, Speech Sound Disorders, Child Speech Processing, Korean Phonology, Speaker Diarization, Self-Supervised Learning
\end{IEEEkeywords}

\section{Introduction}
\label{sec:intro}
Speech sound disorders (SSD), including articulation and
phonological impairments are among the most prevalent communication
difficulties in early childhood, accounting for approximately 44\% of
pediatric communication disorders in South Korea~\cite{choi2019ssd}.
Early identification and targeted intervention are essential, as untreated delays
frequently persist into school age and carry long-term
risks, similar to those associated with many other developmental delays~\cite{eadie2015ssd}~\cite{donthireddy2022orthotic}.
Unfortunately, access to qualified speech-language pathologists (SLPs) remains
constrained by cost, geographic availability, and the time demands of
in-person evaluation, where virtual and automated screening tools
could substantially help.

In South Korea, the Urimal Test of Articulation and Phonology (UTAP) is the
primary standardized protocol for SSD assessment, but its manual scoring is
time-intensive and subject to examiner variability.
To address this,
 23 multidisciplinary experts (12 physicists and 11 SLPs) designed the Hi-DongDong digital assessment tool using a
three-round Delphi process, producing 35 articulation target words.
Hi-DongDong has been validated against both UTAP editions on 931 children aged
2--13 years, demonstrating excellent inter-rater reliability (intraclass
correlation coefficient ICC\,$\geq$\,0.98)~\cite{lee2025hidongdong}.
Our work extends Hi-DongDong's validated word list into an automated end-to-end pipeline for naturalistic home recording settings.

Most existing automated speech evaluation models are developed for
English-speaking adults under controlled acoustic conditions, and do not
generalize to other languages, younger age groups, and/or naturalistic recording
environments~\cite{li2022english}\cite{metallinou2014dnn}.
Korean presents unique challenges due to its fundamental phonological structure, which differs from English in consonant contrasts, vowel inventory, rhythmic timing, and syllable constraints. These differences directly influence the manifestation of pronunciation errors in developing toddler speech evaluation systems.

In this study, the authors focus on Korean children aged 2 to 5 years who are still pre-literate. The evaluation employs a prompted-repetition protocol, where a caregiver models a target word, and the child attempts to repeat it. This process generates multi-speaker recordings that necessitate speaker separation before conducting pronunciation analysis.

The speaker diarization step presents a significant challenge due to the fact that most caregivers are young women who adopt an exaggerated high-pitched speaking style known as aegyo,  a Korean term for acting cute, charming, or baby-like, commonly used to express affection and soften situations. The acoustic profile of aegyo closely resembles that of toddler speech, which complicates conventional speaker diarization approaches \cite{moon2025aegyo}.

\begin{figure}[t]
  \centering
\scalebox{0.9}{
\begin{tikzpicture}[
    node distance = 4mm and 6mm,
    box/.style = {
        rectangle, rounded corners=3pt, draw=black!70, fill=blue!10,
        minimum width=26mm, minimum height=8mm,
        font=\small, align=center, text width=26mm
    },
    decision/.style = {
        diamond, aspect=2, draw=black!70, fill=orange!20,
        font=\small\itshape, align=center, text width=18mm, inner sep=1pt
    },
    output/.style = {
        rectangle, rounded corners=3pt, draw=black!70, fill=green!15,
        minimum width=26mm, minimum height=8mm,
        font=\small\bfseries, align=center, text width=26mm
    },
    arr/.style = {-Stealth, thick, black!70},
    label/.style = {font=\footnotesize\itshape, text=black!60}
]

\node[box] (wav) {\texttt{.wav} recording\\(caregiver + child)};

\node[box, below=of wav] (diar)
      {Speaker Diarization\\(NeMo SortFormer)};

\node[box, below=of diar] (seg)
      {Child Speech\\Segments (word-level)};

\node[box, below=of seg] (ssl)
      {SSL Feature Extraction\\(WavLM / HuBERT)};

\node[box, below=of ssl] (pool)
      {Pooling\\(Within Model Ensemble)};

\node[box, below=of pool] (clf)
      {Logistic Regression Classifier\\(L2 regulation)};

\node[output, below left=8mm and 6mm of clf]  (cons) {Consonant\\Correctness};
\node[output, below right=8mm and 6mm of clf] (vow)  {Vowel\\Correctness};

\draw[arr] (wav)  -- (diar);
\draw[arr] (diar) -- (seg)
      node[label, right, midway] {select child turns};
\draw[arr] (seg)  -- (ssl);
\draw[arr] (ssl)  -- (pool);
\draw[arr] (pool) -- (clf);
\draw[arr] (clf.south) -- ++(0,-4mm) -| (cons.north);
\draw[arr] (clf.south) -- ++(0,-4mm) -| (vow.north);

\node[label, right=2mm of diar]  {Sec.~III-A};
\node[label, right=2mm of ssl]   {Sec.~III-B};
\node[label, right=2mm of pool]  {Sec.~III-B};
\node[label, right=2mm of clf]   {Sec.~III-B};

\end{tikzpicture}
}
  \caption{End-to-end automated pronunciation evaluation pipeline.
           A caregiver--child \texttt{.wav} recording is first segmented into
           speaker turns. Child word segments are then
           encoded by SSL and scored by logistic regression.}
  \label{fig:pipeline}
\end{figure}
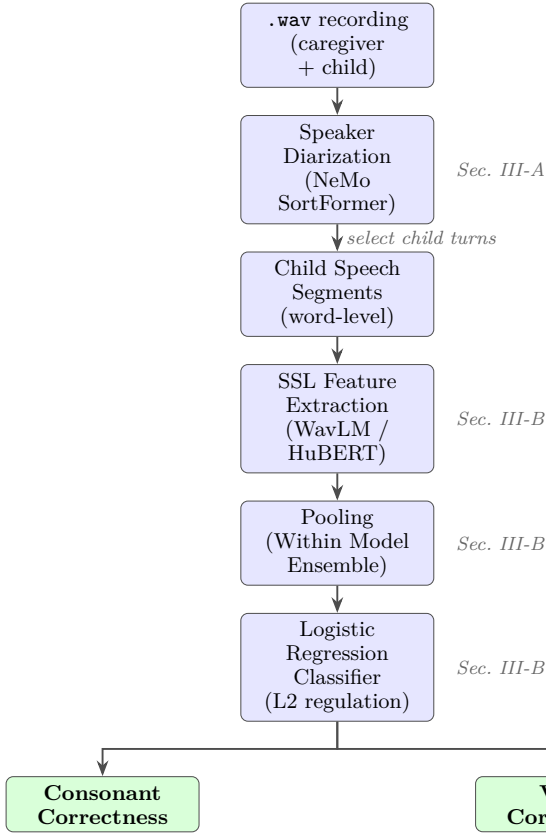

Figure~\ref{fig:pipeline} illustrates the implemented automated pipeline.
A raw \texttt{.wav} recording first passes through speaker diarization
(Section~\ref{subsec:diarization}), which isolates child speech segments
word-by-word.
Each segment is then encoded by a frozen Self-Supervised Learning (SSL) backbone
(Section~\ref{subsec:ssl}), and a lightweight logistic regression classifier
produces binary consonant and vowel correctness.

This paper presents three contributions:
\begin{enumerate}
  \item A novel IRB-approved dataset of Korean toddler speech with word-level
        binary correctness labels for consonants and vowels, annotated by three
        independent trained reviewers.
  \item A systematic benchmark of three representative diarization systems on
        Korean toddler--caregiver recordings.
  \item An automated pronunciation scoring pipeline using pre-trained
        self-supervised speech representations, with a cross-model ensemble
        achieving 0.782 mean balanced accuracy on held-out speakers.
\end{enumerate}

The remainder of this paper is organized as follows.
Section~\ref{sec:related} reviews related work.
Section~\ref{sec:system} describes the system design.
Section~\ref{sec:results} presents the dataset and experimental results.
Section~\ref{sec:conclusion} provides conclusions and discussion.

\section{Related Work}
\label{sec:related}

\subsection{Automated Speech and Pronunciation Assessment}

Early approaches of automated pronunciation evaluation used Hidden Markov Model (HMM)-based forced alignment to compute phoneme-level confidence scores as a
proxy for correctness~\cite{rabiner1989hmm}.
More recent work has applied deep neural networks (DNNs) with substantially
improved accuracy for adult English speakers~\cite{li2022english}.
For child speech specifically, Metallinou
showed that DNN-based automatic speech recognition (ASR) systems using
rectified linear units substantially outperformed HMM baselines for English
language learner children, achieving 31\% relative word error rate reduction ~\cite{metallinou2014dnn}.

The most closely related work to ours applies ASR to diagnose the pronunciation of
speech sound disorders in Korean children~\cite{korean2024asr}.
That system operates on children aged 30--89 months using a standardized
same-device setup with no caregiver-modeled input.
Our study targets a younger range (2--5 years), operates under uncontrolled
multi-device home conditions, and explicitly addresses the caregiver--child
diarization problem.

\subsection{Korean Phonology and Acoustic Challenges}

Korean and English differ across several phonological dimensions relevant to
pronunciation evaluation.
In terms of rhythm, English is \emph{stress-timed} --- stressed syllables recur
at roughly equal intervals and unstressed vowels reduce toward a schwa.
Korean is \emph{syllable-timed} --- each syllable receives approximately equal
duration, producing a more uniform rhythmic pattern~\cite{lee2005syllable}.

The most distinctive feature is Korean's
three-way stop consonant contrast.
English distinguishes stops primarily by voicing (vocal fold vibration,
e.g.,~/p/ vs.\ /b/).
Korean instead uses three laryngeal categories ---
\emph{plain (lax)},
\emph{tense (fortis)}, and
\emph{aspirated} ---
differentiated by voice onset time and laryngeal tension~\cite{cho2002korean}.
This three-way system has no direct English equivalent and is not captured by
models trained solely on English speech.

Additional features relevant to acquisition include the Korean liquid
\textit{ㄹ}, which surfaces as a flap /r/ between vowels
and a lateral /l/ at syllable edges~\cite{iverson1994liquid}, and
\emph{coda neutralization} --- multiple consonants merging and being unreleased
at syllable boundaries~\cite{kimrenaud1974}.

A culturally specific acoustic challenge is \emph{aegyo} (애교), a
communicative style characterized by exaggerated cuteness and childish behaviors, commonly adopted by Koreans females.
Acoustic analyses~\cite{moon2025aegyo} characterize aegyo by significantly
elevated mean fundamental frequency (F$_0$) and maximum F$_0$, greater F$_0$
variability, longer vowel durations, and slower speech rate.
These features overlap substantially with the acoustic profile of toddler
speech: high F$_0$ from shorter vocal tracts, wide pitch variability, and
developmental timing irregularities~\cite{korean2023child}.
Conventional diarization models trained on adult English speech are not designed
to handle this confound.

\subsection{Speaker Diarization}

Speaker diarization is the process of partitioning an audio recording into
segments labeled by speaker identity --- answering ``who spoke
when''~\cite{tranter2006diar}.
While many state-of-the-art systems demonstrated impressive performance on standard English benchmarks, they had not been evaluated in the context of Korean toddler and caregiver speech prior to this research.

Children’s speech presents challenges for speaker diarization because it differs from adult speech, which most diarization models are trained on.  Specifically, children’s speech has higher fundamental frequencies due to shorter vocal tracts, higher formant frequencies, and greater acoustic variability that stabilizes over time~\cite{korean2023child}.  These features make it particularly challenging to separate children’s speech from the co-occurring adult female voices, especially in Korean-speaking environments.

\section{System Design}
\label{sec:system}

\subsection{Speaker Diarization}
\label{subsec:diarization}
\subsubsection{Pre-trained Speaker Diarization Models}
We applied three pre-trained speaker diarization models, including Pyannote,
SpeechBrain, and NeMo with hyperparameter tuning.
For all three models, we set the maximum speaker count to four, accounting for
child, caregiver, and possible additional acoustic sources such as pets or
background speakers.
\begin{itemize}
\item Pyannote combines a neural segmentation
model with speaker embedding and clustering.
It is pre-trained on multilingual corpora ~\cite{bredin2023pyannote}.

\item SpeechBrain implements diarization via
x-vector speaker embeddings and spectral clustering.
x-vectors are fixed-dimensional representations of speaker identity produced by
a time-delay neural network (TDNN) trained with a statistics pooling layer over
frame-level features~\cite{snyder2018xvectors}.
Spectral clustering then groups the resulting embeddings into speaker clusters.
The model is trained primarily on English corpora ~\cite{ravanelli2021speechbrain}.

\item NeMo SortFormer is an encoder-based transformer that solves the permutation problem inherent in multi-speaker settings via an arrival time sorting (ATS) mechanism ~\cite{park2024sortformer}.
Speaker tokens are sorted by order of first appearance, and a
\emph{sort loss} ($\mathcal{L}_\text{Sort}$) trains the model to learn temporal
ordering jointly with speaker labels.
The loss function is defined as:
\begin{equation}
  \mathcal{L}_\text{Sort}\!\left(Y,\, f_\theta(X)\right)
    = \mathcal{L}_\text{BCE}\!\left(Y_\eta,\, f_\theta(X)\right)
  \label{eq:sort_loss}
\end{equation}
\noindent where $\mathcal{L}_\text{BCE} = -(y\log p + (1-y)\log(1-p))$ is
binary cross-entropy, $f_\theta(X)$ is the model output, $y$ is the true
speaker label, $p$ is the predicted speaker probability, and $Y_\eta$ is the
label vector sorted by arrival time using sorting function $\eta$.
\end{itemize}
\subsubsection{Evaluation Metric}
Diarization Error Rate ($DER$) quantifies the proportion of total reference speech time that is incorrectly attributed. It aggregates missed speech ($MS$), false alarm ($FA$), and speaker confusion ($SC$) errors~\cite{tranter2006diar}. DER serves as a metric to assess the performance of our model in diarizing speech from Korean-speaking children and caregivers.

\begin{equation}
  DER = \frac{MS + FA + SC}{Total\, Reference\, Speech\, Time}
  \label{eq:der}
\end{equation}

\subsection{Self-Supervised Speech Evaluation Model}
\label{subsec:ssl}
\subsubsection{Pre-trained Self-Supervised Speech Evaluation Models}
Self-supervised learning (SSL) for speech refers to a paradigm in
which large transformer-based models are pre-trained on massive amounts of
unlabeled audio using tasks that require no human annotation, such as masked
frame prediction, where the model must reconstruct hidden portions of the input
waveform.
During this pre-training phase, the model learns rich acoustic representations
that encode fine-grained phonetic, prosodic, and speaker information from raw
waveforms alone~\cite{baevski2020wav2vec2}\cite{chen2022wavlm}\cite{hsu2021hubert}.

In this study, we employ the SSL encoder as a frozen feature extractor. Each word-level audio segment undergoes processing through the encoder to generate a sequence of frame-level embeddings. These embeddings are subsequently aggregated into a single vector and fed into a supervised logistic regression classifier. This classifier is trained on our binary consonant and vowel correctness labels. The two-stage design enables the model to leverage the rich acoustic representations while simultaneously training the downstream classifier using our limited annotated dataset.

We evaluate three backbone models:
\begin{itemize}
  \item {wav2vec2-large-xlsr-korean}~\cite{kresnik2021xlsr}
        : A multilingual wav2vec~\cite{baevski2020wav2vec2} model fine-tuned specifically on Korean speech data. 
        This is the only Korean-specific model in our comparison.
  \item {HuBERT-large}~\cite{hsu2021hubert}: Hidden Unit BERT ---
        a large SSL model using masked discrete unit prediction, pre-trained
        on 960\,hours of English LibriSpeech~\cite{panayotov2015librispeech} then fine-tuned for ASR.

  \item {WavLM-large}~\cite{chen2022wavlm}: Pre-trained on 94,000 hours of diverse English audio with a masked speech denoising objective, WavLM predicts clean speech under noisy conditions. This objective explicitly encourages the encoding of fine-grained phonetic content, making it particularly well-suited for pronunciation-level tasks.
\end{itemize}

\subsubsection{Pooling Strategies}

Each word segment produces a variable-length sequence of frame embeddings
(typically 768--1{,}024\,dimensions per frame).
We evaluated three strategies to aggregate these into a fixed-size vector:
\begin{itemize}
  \item Mean pooling: simple average across all frames
        from the final transformer layer.
        Captures average phonetic content but discards temporal variability.

  \item Attention pooling: a small dedicated sub-network sub-network to determine and assign a scalar attention score to each frame via softmax, then computes the weighted
  sum. Allows the model to focus on the most informative frames.

  \item Statistics pooling: concatenates the
        frame-wise mean and standard deviation vectors.
        Captures both average acoustic content and temporal variability within
        the word.

  \item Multi-layer fixed-weight pooling: averages representations across a
        fixed range of transformer layers, exploiting that different layers encode different levels
        of phonetic abstraction.

\end{itemize}

\subsubsection{Classifier}

The pooled embedding is fed into a logistic regression classifier with
$L2$ regularization.
$L2$ regularization adds a penalty proportional to the squared magnitude of the
weight vector $\mathbf{w}$ (the logistic regression coefficient vector) to the
loss function.
\begin{equation}
  \mathcal{L} = \mathcal{L}_\text{CE} + \frac{1}{C}\,\|\mathbf{w}\|^2
  \label{eq:l2}
\end{equation}
\noindent where $\mathcal{L}_\text{CE}$ is cross-entropy loss,
$\mathbf{w} \in \mathbb{R}^d$ is the logistic regression weight vector
($d$ = embedding dimensionality), and $C$ controls the regularization strength. 
We use $C = 1.0$, providing moderate regularization appropriate for our feature
dimensionality ($d \approx 2{,}048$ for statistics pooling of WavLM-large).
Separate classifiers are trained for consonant and vowel tasks.

\subsubsection{Evaluation Metric}

In this study, balanced accuracy (BA) is used as a primary metric for classification.
\begin{equation}
\begin{split}
  \text{BA} &= \frac{1}{2}\!\left(\frac{TP}{TP+FN} + \frac{TN}{TN+FP}\right)
\end{split}
\label{eq:ba}
\end{equation}

\noindent where $TP$ and $TN$ are true positives and negatives, whereas $FN$ and $FP$ are false positives and negatives.
A classifier that always predicts the majority class achieves
$\text{BA} = 0.5$ regardless of class distribution, making BA the appropriate
metric for our class-imbalanced dataset.

\section{Results}
\label{sec:results}

\subsection{Dataset}
\label{subsec:dataset}

\subsubsection{Data Collection}
We collected 53 recordings from Korean-speaking children aged between 2 and 5 years old.
Recordings were collected by caregivers at home or by teachers at childcare
facilities, following the Hi-DongDong assessment protocol~\cite{lee2025hidongdong}.
The caregiver models each target word, and the child repeats.
Participants used their own personal devices (smartphones, tablets, laptops),
producing variation in microphone sensitivity and channel quality.
Background noise sources include ambient noise, background conversation, and
incidental questioning.
Among the 53 recordings, 9 were made without audible caregiver speech; the
remaining 44 include caregiver model utterances.
Of the 44 caregivers, 42 were female, and 2 were male.
The recordings had a mean duration of 89.10 seconds, a standard deviation of 31.18 seconds, a minimum duration of 30.30 seconds, a maximum duration of 173.97 seconds, a median duration of 85.60 seconds, and a total corpus of 60.9 minutes.

\subsubsection{Target Word List}
Table~\ref{tab:wordlist} lists the 35 Hi-DongDong target words with English
translations and phonological features.
Of the 35 words, 22 were evaluated for both consonant and vowel correctness;
the remaining 13 are consonant-only evaluations (marked ``N'' in the ``Vowel''
column), as they do not contain phonologically salient vowel targets in the
Hi-DongDong framework.

\begin{table}[t]
  \centering
  \caption{Target word list. ``Vowel?''\ :
           Evaluated for vowel correctness. \\CV = Cons.+ Vowel, CVC = Cons. + Vowel + Cons.}
  \label{tab:wordlist}
  \scriptsize
  \begin{tabular}{clllc}
    \toprule
    \textbf{\#} & \textbf{Korean} & \textbf{English} &
      \textbf{Phonological Feature} & \textbf{Vowel?} \\
    \midrule
     1 & 타조   & Ostrich    & Aspirated stop \ipa{/tʰ/}              & Y \\
     2 & 냄비   & Pot        & Nasal cluster \ipa{/nm/}               & N \\
     3 & 눈썹   & Eyebrow    & Tense fricative \ipa{/s͈/} + coda      & Y \\
     4 & 딸기   & Strawberry & Tense stop \ipa{/t͈/}                  & N \\
     5 & 국자   & Ladle      & Velar stop \ipa{/k/} + affricate       & Y \\
     6 & 다리   & Bridge     & Liquid \ipa{/ɾ/}                       & N \\
     7 & 자동차 & Car        & Three-syllable affricate chain         & N \\
     8 & 씨앗   & Seed       & Tense sibilant \ipa{/s͈/} + vowel      & Y \\
     9 & 바나나 & Banana     & Bilabial \ipa{/b/}, three-syllable     & Y \\
    10 & 레몬   & Lemon      & Liquid onset \ipa{/ɾ/}                 & N \\
    11 & 코끼리 & Elephant   & Aspirated \ipa{/kʰ/} + tense \ipa{/k͈/} & Y \\
    12 & 곰     & Bear       & Simple CVC \ipa{/ko.m/}                & Y \\
    13 & 사탕   & Candy      & Aspirated stop \ipa{/tʰ/}              & N \\
    14 & 버섯   & Mushroom   & Coda neutralization \ipa{/t/}          & Y \\
    15 & 펭귄   & Penguin    & Final nasal + \ipa{/w/}-glide          & Y \\
    16 & 마이크 & Microphone & Loanword, mid vowels                   & Y \\
    17 & 연필   & Pencil     & Final liquid \ipa{/l/}                 & N \\
    18 & 까치   & Magpie     & Tense stop \ipa{/k͈/}                  & N \\
    19 & 빵     & Bread      & Tense bilabial \ipa{/p͈/}              & N \\
    20 & 아빠   & Dad        & Tense bilabial \ipa{/p͈/} (easiest)    & N \\
    21 & 책상   & Desk       & Aspirated \ipa{/tɕʰ/} + tense \ipa{/s͈/} & Y \\
    22 & 찌개   & Stew       & Tense affricate \ipa{/t͈ɕ/}            & N \\
    23 & 오뚜기 & Tumbler    & Tense medial \ipa{/t͈/} + vowels       & Y \\
    24 & 하마   & Hippo      & Simple CV                              & N \\
    25 & 구급차 & Ambulance  & Four-syllable, complex coda            & Y \\
    26 & 수박   & Watermelon & Final stop \ipa{/k/}                   & N \\
    27 & 과자   & Snack      & \ipa{/wa/} diphthong                   & Y \\
    28 & 돼지   & Pig        & \ipa{/we/} diphthong                   & Y \\
    29 & 요거트 & Yogurt     & Loanword, \ipa{/ɯ/} vowel              & Y \\
    30 & 귀     & Ear        & \ipa{/wi/} diphthong                   & Y \\
    31 & 여우   & Fox        & Vowel sequence \ipa{/jʌ.u/}            & Y \\
    32 & 우유   & Milk       & Vowel sequence \ipa{/u.ju/}            & Y \\
    33 & 야구   & Baseball   & Glide onset \ipa{/j/}                  & Y \\
    34 & 원숭이 & Monkey     & Complex nucleus \ipa{/wʌ/}             & Y \\
    35 & 의자   & Chair      & \ipa{/ɰi/} vowel (rare in Korean)      & Y \\
    \bottomrule
  \end{tabular}
\end{table}

\subsubsection{Annotation}
From the original recordings of 53 individuals, we narrowed down to 34 subjects whose recordings featured multiple speakers, and the child’s speech was evaluated using more than 10 words.

For diarization, a native Korean speaker annotated the start and end timestamps of each word in milliseconds. However, due to variations in recording device settings and the study’s nature, where many speakers overlap, there may be errors in the diarization labeling process.

For classification, three trained reviewers independently evaluated each target word spoken by the child, following Hi-DongDong’s protocols.
For each word, reviewers provided two binary judgments --- consonant correctness
and vowel correctness.
Reviewers also annotated the target phoneme(s) being evaluated and documented
the child's actual production of consonants, providing a phonetic error log
that can support future fine-grained error analysis.
Final labels were determined by quorum voting for this study. A word is marked as
incorrect if at least two of three reviewers judge it so.


\subsubsection{Classification Label Distribution}
The labeled dataset contains 1{,}190 consonant labels and 748 vowel labels
across 34 subjects.
Overall consonant failure rates are 24.8\% (295/1{,}190) and vowel failure rates
are 13.5\% (101/748), indicating that the majority class is ``correct'' for
both tasks --- a class imbalance that makes balanced accuracy the appropriate
evaluation metric.
Table~\ref{tab:age} shows failure rates broken down by age group.

\begin{table}[t]
  \centering
  \caption{Consonant and vowel failure rates by age group (34 annotated
           subjects). Failure rate = proportion of labels marked as incorrect by
           quorum voting. Younger children show substantially higher error
           rates, consistent with phonological acquisition milestones.}
  \label{tab:age}
  \begin{tabular}{lcccc}
    \toprule
    \textbf{Age Group} & \textbf{$N$} &
      \textbf{Cons.\ Failure} & \textbf{Vowel Failure} \\
    \midrule
    2 years  & 16 & 38.9\% & 21.6\% \\
    3 years  &  7 & 15.9\% &  6.5\% \\
    4 years  &  8 &  9.6\% &  6.2\% \\
    5 years  &  3 & 10.5\% &  6.1\% \\
    \bottomrule
  \end{tabular}
\end{table}


\subsection{Diarization Results}
\label{subsec:diar_results}

Speaker count accuracy was computed across all 53 recordings.
DER was computed on the 34 recordings confirmed to contain exactly two
speakers each speaking more than 10 words, with a collar of 0.16\,seconds at
speaker boundaries not to penalize transitions within this window.

\begin{table}[t]
  \centering
  \caption{Diarization model comparison.
           Speaker count accuracy computed on all 53 recordings.
           DER computed on 34 two-speaker recordings.}
  \label{tab:diar}
  \begin{tabular}{lcc}
    \toprule
    \textbf{Model} & \textbf{Spk.\ Count Acc.} & \textbf{Overall DER} \\
    \midrule
    NeMo SortFormer   & \textbf{88.68\%} & \textbf{33.04\%} \\
    Pyannote.audio    & 62.26\% & 154.36\% \\
    SpeechBrain       & 43.40\% & 136.21\% \\
    \bottomrule
  \end{tabular}
\end{table}

NeMo SortFormer substantially outperforms both alternatives, achieving 88.68\%
speaker count accuracy and 33.04\% DER.
The high DER values for Pyannote and SpeechBrain reflect systematic failure to
separate child and caregiver segments in this corpus, attributable to the
acoustic similarity between children and caregivers.

While the 33.04\% DER represents an advanced benchmark for this demographic, the remaining error, mainly due to overlapping high-pitched aegyo sounds, inevitably introduces noise into the downstream classification process. 
\subsection{Classification Results}
\label{subsec:class_results}

Each word spoken by a toddler was extracted from the word segment timestamps generated by the NeMo diarization pipeline. The data was divided into an 80/20 speaker-level split, resulting in 27 speakers for training and 7 speakers for testing. Each word token from a particular child speaker was assigned exclusively to one partition.

\paragraph {Pooling Strategy Comparison}

Tables~\ref{tab:wavlm_pooling} and~\ref{tab:hubert_pooling} compare pooling strategies for WavLM-large and HuBERT-large, respectively. Since no single strategy is universally optimal, a within-model ensemble combining the two strongest strategies via soft voting (averaging their predicted error probabilities before thresholding at $0.5$) was employed. This approach allows partial confidence from each strategy to cancel opposing errors. For both WavLM-large and HuBERT-large, an ensemble of statistics and multi-layer weight pooling strategies was used as the within-model ensemble.
\begin{table}[htbp]
  \centering
  \caption{WavLM-large: balanced accuracy (BA) by pooling strategy.
           }
  \label{tab:wavlm_pooling}
  \begin{tabular}{lccc}
    \toprule
    \textbf{Pooling Strategy} &
      \textbf{Cons.\ BA} & \textbf{Vowel BA} & \textbf{Mean BA} \\
    \midrule
    Mean         & 0.632 & 0.703 & 0.668 \\
    Attention          & 0.646 & 0.777 & 0.712 \\
    Statistics       & 0.699 & 0.831 & 0.765 \\
    \makecell{Multi-layer Fixed Weights\\ (cons:1--6 / vow:7--12)}  & 0.502 & 0.849 & 0.676 \\
    \midrule
    Within-model ensemble      & 0.693 & 0.845 & 0.769 \\
    \bottomrule
  \end{tabular}
\end{table}

\vspace{-1em}
\begin{table}[htbp]
  \centering
  \caption{HuBERT-large: balanced accuracy (BA) by pooling strategy.}
  \label{tab:hubert_pooling}
  \begin{tabular}{lccc}
    \toprule
    \textbf{Pooling Strategy} &
      \textbf{Cons.\ BA} & \textbf{Vowel BA} & \textbf{Mean BA} \\
    \midrule
    Mean           & 0.667 & 0.633 & 0.650 \\
    Attention            & 0.641 & 0.721 & 0.681 \\
    Statistics     & 0.708 & 0.669 & 0.689\\
    \makecell{Multi-layer Fixed Weights\\ (cons:1--11 / vow:12--24)} & 0.708 & 0.651 & 0.680 \\
    \midrule
    Within-model ensemble       & 0.720 & 0.651 & 0.685 \\
    \bottomrule
  \end{tabular}
\end{table}
For WavLM-large, statistics pooling is the dominant method for both consonant and vowel tasks, achieving a higher accuracy of 0.699 and 0.831, respectively. In contrast, for HuBERT-large, attention pooling performs better on vowel evaluation tasks, while statistics pooling demonstrates higher mean balanced accuracy.

\paragraph {Cross-Model Comparison}

\begin{table}[t]
  \centering
  \caption{Speech evaluation model comparison.
           Balanced accuracy (BA): 
           Statistics pooling is used for all single-model entries.
           $^\ast$Cross-model Ensemble uses HuBERT-large for consonant
           prediction (BA\,=\,0.720) and WavLM-large for vowel prediction
           (BA\,=\,0.845) with the ensemble pooling method.}
  \label{tab:class}
  \begin{tabular}{llccc}
    \toprule
    \textbf{Model} &
      \textbf{Cons.\ BA} & \textbf{Vowel BA} & \textbf{Mean BA} \\
    \midrule
    wav2vec2-XLSR-Korean~\cite{kresnik2021xlsr}
     & 0.583 & 0.480 & 0.531 \\
    HuBERT-large~\cite{hsu2021hubert}
      & 0.708 & 0.669 & 0.689 \\
    WavLM-large~\cite{chen2022wavlm}
      & 0.699 & 0.831 & 0.765 \\
    \bottomrule
  \end{tabular}
\end{table}

wav2vec2-XLSR-Korean~\cite{kresnik2021xlsr} --- the only
Korean-specific model --- achieves 0.583 consonant BA and 0.480 vowel BA.
Despite its Korean pre-training, it underperforms WavLM-large and HuBERT-large on both tasks, suggesting that its Korean ASR fine-tuning optimizes representations for Korean
phoneme recognition, but less so for the fine-grained acoustic differences
in toddlers.

WavLM-large~\cite{chen2022wavlm} achieves the highest vowel result (0.845 BA) in single-model evaluation using within-model ensemble pooling. Its denoising pre-training objective, which involves predicting clean speech under noisy conditions, seems to preserve fine-grained spectral features that are crucial for distinguishing vowel qualities. On the other hand, HuBERT-large~\cite{hsu2021hubert} leads in consonant recognition (0.720 BA). This result aligns with HuBERT’s discrete unit prediction objective, which captures articulatory patterns that are essential for accurate consonant production.

The cross-model ensemble routes each task to its most powerful backbone, ensuring that the most effective model for each task is utilized. By assigning HuBERT-large for consonant prediction (0.720 BA) and WavLM-large for vowel prediction (0.845 BA), the ensemble achieves an overall BA of 0.782. This task-routing strategy effectively avoids the limitations of simple model averaging and simultaneously achieves the best performance of both models.



\section{Conclusion}
\label{sec:conclusion}

This work introduced an automated end-to-end pipeline for pronunciation
evaluation of Korean toddler speech, combining neural speaker diarization with
self-supervised speech representations.
Our experiments suggest that SSL models pre-trained on adult English speech can transfer meaningfully to Korean toddler pronunciation assessment, achieving a mean balanced accuracy of 0.782 with a cross-model
ensemble.

Among the three diarization systems evaluated, NeMo SortFormer achieved the lowest DER (33.04\%), substantially outperforming Pyannote.audio and SpeechBrain. Its arrival-time-sorted transformer architecture appears particularly effective at resolving the acoustic confound between young female caregivers exhibiting aegyo and toddler speech, a challenge absent from standard English diarization benchmarks. Critically, by isolating child word segments automatically, the pipeline eliminates the manual segmentation step that would otherwise precede self-supervised speech evaluation, substantially reducing the time and effort required for large-scale assessment. The residual DER will introduce some segmentation errors that propagate downstream; future work will quantify this effect by comparing evaluation accuracy on automatically diarized versus manually segmented inputs.

The complementarity of WavLM and HuBERT across tasks, WavLM for
vowels, HuBERT for consonants, is consistent with the different emphases of their pre-training objectives ~\cite{chen2022wavlm}\cite{hsu2021hubert}.
WavLM's denoising training may encourage encoding of spectral envelope features relevant to vowel quality, while HuBERT's discrete unit prediction likely captures categorical articulatory distinctions relevant to consonant accuracy.
These findings suggest that a \textit{divide-and-conquer} ensemble strategy
provides a robust framework for low-resource pediatric speech assessment
without domain-specific fine-tuning.


The authors plan to expand the labeled dataset, especially for the 4–5 year age group where class imbalance is most severe. The expanded dataset will enable leave-one-speaker-out cross-validation for more reliable performance estimates. Additionally, the authors plan to develop a mobile application to streamline data collection, which would simultaneously reduce diarization errors by guiding recording conditions and accelerate dataset expansion. Further directions include end-to-end fine-tuning of the SSL backbone on Korean child speech and phoneme-level alignment for more granular error localization.

\section*{Acknowledgments}
This work was supported by the National IT Industry Promotion Agency (NIPA)
grant funded by the Korean Government (MSIT) (No.\ PJT-25-040355).
This study was also supported by Grant No.\ 14-2025-0015 from the SNUBH
Research Fund.

\bibliographystyle{IEEEtran}
\bibliography{references}

\end{document}